\documentclass[12pt]{article}
\usepackage{amsmath}
\usepackage{amsthm}
\usepackage{amssymb}
\def\bea{\begin{eqnarray}}
\def\beq{\begin{equation}}
\def\eea{\end{eqnarray}}
\def\eeq{\end{equation}}
\newcommand{\im}{{\rm i}}

\newcommand{\ri}{{\rm i}}
\begin{document}
\begin{center}
{\bf \Large  Classical equations of motion and algebras of quantum observables}
\footnote{This work was supported in part by the Russian Fund of Basic Researches (grant No 09 - 01 - 00504)}
\end{center}

\begin{center}
{\bf M.A. Sokolov}
\end{center}
\begin{center}
{\it Department of Physics, St Petersburg Institute of Machine Building,
\\
Poliustrovskii 14, St Petersburg, Russia}
\\
{mas@ms3450.spb.edu,  masokolov@gmail.com}
\end{center}

In this work simple and effective quantization procedure of classical dynamical systems is proposed and illustrated by a number of examples. The procedure is based entirely on differential equations which describe time evolution of systems.

\section{Introduction}

The aim of this work is to propose and illustrate by a number of examples a simple method of quantization of classical finite-dimensional dynamical systems. Recall that a quantization in a broad sense is a transition from classical description of a dynamical system to its quantum description. In the early days of quantum mechanics the quantization could be resumed as selecting some solutions of classical dynamical equations according to Bohr-Sommerfeld rules. Subsequently quantization was understood in a new fashion as replacing of classical observables by operators acting in Hilbert space of states. The prescription for such replacement was formulated in the late twenties of the last century and was called canonical quantization.

According to the canonical quantization procedure  \cite{D}, \cite{vN} classical coordinates and con\-ju\-gated momenta are replaced by Hermitian operators $\hat q^i,\hat p_i$ which subject to the following canonical commutation relations

\begin{equation}
\label{qz1-ccr}%
[\hat q^i,\hat q^j] = 0,\quad %
[\hat p_i,\hat p_j] = 0,\quad %
[\hat q^i,\hat p_j] = \im\hbar\,\delta ^i_{j}\,\hat{\mathbb{I}}%
\quad  i,j = 1,\,...\, n.
\end{equation}
Here $\delta ^i_{j}$ is the Kronecker delta and $\hat{\mathbb{I}}$
is the identity operator which, as a rule, will be omitted in what follows.
Other classical observables  $f(q^i,p_j)$ are replaced by quantum
observables $\hat f = f(\hat q^i,\hat p_j)$. Further, in the context of the procedure quantum dynamical equations originate from the classical Hamiltonian equations

\begin{equation}
\label{qz1-2}%
\frac{{\rm d}q^i}{{\rm d}t} = \{q^i,H\},\quad %
\frac{{\rm d}p_i}{{\rm d}t} = \{p_i,H\},\quad %
i = 1,\,...\, n %
\end{equation}
which are written in Poisson bracket form. Recall that the canonical brackets of two arbitrary functions $f$ and $g$ are defined by the following formula

\begin{equation}
\label{qz1-cpb}%
\{f,g\}=
\sum _{i=1}^{n}\left(\frac{\partial f}{\partial q^i}\frac{\partial g}{\partial p_i} - %
\frac{\partial g}{\partial q^i}\frac{\partial f}{\partial p_i}\right). %
\end{equation}
The replacement of the brackets (\ref{qz1-cpb}) by the commutator

\begin{equation}
\label{qz1-3}%
\{f,g\} \longrightarrow \frac{1}{\im\hbar}[\hat f,\hat g]
\end{equation}
in the classical equations of motion (\ref{qz1-2}) gives the quantum Heisenberg equations

\begin{equation}
\label{qz1-4}%
\frac{{\rm d}\hat q^i}{{\rm d}t} = \frac{1}{\im\hbar}[\hat q^i,\hat H],\quad %
\frac{{\rm d}\hat p_i}{{\rm d}t} = \frac{1}{\im\hbar}[\hat p_i,\hat H],\quad %
i = 1,\,...\, n %
\end{equation}
where $\hat H = H(\hat q^i,\hat p_j)$ is a quantum Hamiltonian.

The canonical quantization algorithm works perfectly for most Hamiltonian systems, in particular for the systems which Hamiltonian function has the form

\begin{equation}
\label{qz1-5}%
H = \sum \frac{{p_i}^2}{2m_i} + V(q^1, ... ,q^n). %
\end{equation}
However, for dynamical systems with Hamiltonians more complicated than (\ref{qz1-5}), for multi-Hamilto\-nian systems, for systems which not allows global Hamiltonian structure, and in some other cases the canonical quantization often gives indeterminate results or it is non-applicable at all. In addition there is evident dependence of canonical quantization results on coordinate choice.

These reasons have stimulated on the one hand, revision of  the canonical quantization algorithm (for example, geometric quantization \cite{Wdhs}) and on the other hand search new quantiza\-tion methods (path integral quantization \cite{FH}, deformation quantization \cite{DefQuant}, etc.). At present there are a broad spectrum of various quantization schemes and a rich variety of them with the related references are listed in the recent review \cite{STAME}.

Almost all quantization methods imply the existence of globally defined symplectic, Poisson or more rich structures on a phase manifold of a classical system to be quantized. However, certain of the dynamical systems does not allow such a structures or these structures are complicated enough (for example, dissipative systems). In this case a quantization method which is relied solely on classical dynamical equations looks very attractive. Remark, that possibility to deduce commutation relations among coordinates and momentum from equations of motion was discussed long ago in the papers \cite{Wigner}, \cite{Okubo}. Since that time a few works were made in this direction. Here we mention the paper \cite{Taras} where the Weil quantization scheme was employed to obtain commutation relations and quantum dynamical equations for some dissipative systems.

In the present work a simple  quantization algorithm is proposed in which classical equations of motion play crucial role. Namely, for any classical system the algorithm allows to determine simultaneously quantum commutators for initially commutative dynamical variables and quantum corrections to the original classical dynamical equations. As a result we obtain commutation relations for any pair of basis quantum  variables $\hat x ,\hat y,\, ...\,\hat z$ and  equations of motion  in the form
$$%
\hat y\hat x=\hat x\hat y +\sum _{k=1}\hbar ^k f_k(\hat x ,\, ...\,,\hat z)
$$

$$%
\frac{{\rm d}\hat x}{{\rm d}t} = r(\hat x ,\, ...\,\hat z) + \sum
_{k=1}\hbar ^k g_k(\hat x ,\, ...\,\hat z).
$$
In the above relations $\hbar$ is Plank constant, $f_k,\,g_k$ and $r$ are functions of specially ordered quantum variables. These functions can be defined from the conditions that commutation relations are time independent and quantum variables
are Hermitian. The same conditions allow to obtain quantum corrections for integrals of motion if they exist.

Thus, for any classical system the proposed quantization algorithm gives in our disposal an algebra of quantum observables with a set of commutation relations
among its base elements and a system of quantum evolution equations. To investigate the obtained quantum system it is necessary to choose appropriate representation of the quantum observables algebra. This choice essentially depends on form of commutation relations and is beyond the algorithm. So in
the present work we concern only formal algebraic aspects of the quantization procedure and almost entirely ignore other ones.

In the next section the quantization algorithm will be presented in some
details and then illustrated by several well known examples (charged spinless
particle in a magnetic field, anisotropic Euler top, etc.). Some remarks concerning the algorithm will be given in conclusion.

\section{Quantization algorithm}

Consider a finite - dimensional dynamical system which evolution is defined by a system of ordinary differential equations or, in geometric terms, by some vector field ${X}$ (see, for example, \cite{Arnold}). More precisely, let $M$ be a real $n$ - dimensional manifold and ${\cal F}(M)$ the algebra of infinitely differentiable
complex - valued functions on $M$. From the physical point of view $M$ is a phase manifold and ${\cal F}(M)$ is an algebra of observables of this system. Time evolution of any function $f$ from ${\cal F}(M)$ is determined by the equation

\begin{equation}
\label{qz2-1}%
\frac{{\rm d}f}{{\rm d}t} = {X}({f}), \quad f =f(x)\in{\cal F}(M),
\quad x\in M%
\end{equation}
where  ${X}$ is the vector field under consideration. Recall that
any vector field is a derivation of ${\cal F}(M)$. This means that
the linearity condition

$$ {X}(af+bg) = a{X}(f) + b{X}(g)$$
and Leibnitz rule

$${X}(fg) = {X}(f)g + f{X}(g)$$
hold for arbitrary functions $f,g\in {\cal F}(M)$ and complex
numbers $a,b \in \mathbb{C}$. In local coordinates $x^k,\,k=1, ...
,n,$  the field $X$ has the form

\begin{equation}
\label{qz2-2}%
X = \sum _{k=1}^{n}{X}^k(x){\partial}_k, \quad %
{\partial}_k = \frac{\partial}{\partial {x^k}}, \quad %
{X}^k(x)\equiv  X(x^k)\in {\cal F}(M).%
\end{equation}%
Using the equation (\ref{qz2-1}) we can obtain evolution equations
for basis dynamical variables which we identify here with the
coordinate functions $x^k = x^k(x)$

\begin{equation}
\label{qz2-3}%
\frac{{\rm d}x^k}{{\rm d}t} = {X}^k(x).%
\end{equation}
Equations of the form (\ref{qz2-3}) usually derived from experimental data.

Solution of equations (\ref{qz2-3}), at least when the functions
${X}^k(x)$ are analytical, can be represented in the form

\begin{equation}
\label{qz2-3s}%
x^i(t) = x^i(t,x_0^1, ... ,x_0^n)=%
\sum _{k_1,\, ...\, ,k_n\ge 0}s^i_{k_1,\, ...\,
,k_n}(t)(x_0^1)^{k_1} ... (x_0^n)^{k_n}
\end{equation}
where $x_0^i,\,i=1, ... ,n$ are the values of the coordinate functions
at $t=0$ and $s^i_{k_1,\, ...\, ,k_n}(t)$ are real functions of $t$.
This solution defines a one-parameter group of diffeomorphisms of
$M$ 

$$
g_{X}^t\,:\,M\rightarrow M,\quad g_{X}^tx(0)=x(t), \quad x\in M.
$$
Observables which are invariant under diffeomorphisms from this
group are called integrals of motion of a dynamical system and can
be obtained from the equation

\begin{equation}
\label{qz2-4}%
\frac{{\rm d}I(x)}{{\rm d}t}={X}(I)=0,\quad I(x)\in {\cal F}(M). %
\end{equation}

To realize the quantization procedure we firstly perform more or
less formal transition from the algebra of classical observables
${\cal F}(M)$ to a commutative algebra ${\cal F}(\tilde M)$. The
transition is defined by the map $\kappa :x^i\mapsto \tilde x^i$
which associates basis elements $\tilde x^i$ of the algebra ${\cal
F}(\tilde M)$ with the coordinate functions $x^i=x^i(x)\in {\cal
F}(M)$. This map can be extended naturally to the whole algebra
${\cal F}(M)$. Namely, for any function $f(x^1, ... ,x^n)\in {\cal
F}(M)$ we obtain 

\begin{equation}
\label{qz2-20}%
\kappa(f(x^1, ... ,x^n)) = f(\kappa(x^1), ... ,\kappa(x^n))=
f(\tilde x^1, ... ,\tilde x^n)\in {\cal F}(\tilde M).
\end{equation}
The above formula will be quite defined if we add to the set of basis
elements $\tilde M = (\tilde x^1, ... ,\tilde x^n)$ a multiplicative
unit $\tilde{\mathbb{I}}$. So let us put

\begin{equation}
\label{qz2-20a}%
\kappa(a) = a\tilde{\mathbb{I}}\in {\cal F}(\tilde M), \quad \forall
a\in \mathbb{C}.
\end{equation}
Now introduce an involution in ${\cal F}(\tilde M)$ taking into
account reality of coordinate functions $\bar x^i =x^i$ where bar
denote a complex conjugation in ${\cal F}(M)$. Set for any $f\in
{\cal F}(M)$

\begin{equation}
\label{qz2-7h} \kappa({\bar f}) = (\kappa(f))^{*}.
\end{equation}
Here star denote involution in ${\cal F}(\tilde M)$. From this
formula and the reality $x^i$ it follows that the basis elements of
${\cal F}(\tilde M)$ are Hermitian

\begin{equation}
\label{qz2-7m}%
(\tilde x^k)^{*} = {\tilde x}^k.
\end{equation}
Thus ${\cal F}(\tilde M)$ is a commutative unital
$\mathbb{C}$-algebra with involution and its elements are  formal
series in basis elements $\tilde M = (\tilde x^1, ... ,\tilde x^n)$.

To represent dynamic of our system in terms of $\tilde x^i$ define
the vector field $X$ counterpart $\tilde X$  as a derivation in the algebra ${\cal F}(\tilde M)$ which acts on the basis elements ${\tilde x}^k$ according to the formulas

\begin{equation}
\label{qz2-7}%
{\tilde X}({\tilde x}^k)=\widetilde{X(x^k)},
\quad k = 1, ... ,n, \quad {\tilde X}(\mathbb{\tilde I})=0.%
\end{equation}
This relation and the fact that the series
\begin{equation}
\label{qz2-8}%
\tilde x^i(t) = \tilde x^i(t,\tilde x_0^1, ... ,\tilde x_0^n)=%
\sum _{k_1,\, ...\, ,k_n}s^i_{k_1,\, ...\, ,k_n}(t)(\tilde
x_0^1)^{k_1} ... (\tilde x_0^n)^{k_n},\quad i=1, ... ,n, %
\end{equation}
is a formal solution of the equations

\begin{equation}
\label{qz2-6}%
\frac{{\rm d}\tilde x^k}{{\rm d}t} = {\tilde X}(\tilde x^k) %
\end{equation}
if the series (\ref{qz2-3s}) is a solution of the equations
(\ref{qz2-3}) suggest that the equation (\ref{qz2-6}) can be accepted
as a  dynamical equations for basis elements $\tilde x^k$.

Thus, as a result of our preliminary formal step we obtain  (associative)
commutative involutive $\mathbb{C}$ - algebra ${\cal F}(\tilde M)$ with
Hermitian basis elements ${\tilde x}^i$ and fixed derivation $\tilde{X}$ of this algebra which defines the evolution equations (\ref{qz2-6}). In addition if the original classical system allows integrals of motion $I_j,\,j = 1,...,m<n,$ we obtain a set of such integrals defined by the equation

\begin{equation}
\label{qz2-76}%
\tilde{X}(\tilde I_j)=0,\quad j = 1,...,m<n, \quad \tilde I_j\in {\cal F}(\tilde M).
\end{equation}

Now we are ready for the final step of the quantization algorithm, namely for a transition from the commutative algebra ${\cal F}(\tilde M)$ to a noncommutative algebra of quantum observables ${\cal F}(\hat M)$. Let under the transition each basis element $\tilde x^i\in {\cal F}(\tilde M)$ converts in a basis element $\hat x^i\in {\cal F}(\hat M)$ and the unit $\tilde{\mathbb{I}}$ converts in its counterpart $\hat{\mathbb{I}}$. Further,
define  an order of multipliers in any product of the  elements $\hat x^i$. Fix, for example, for any monomial in basis elements $\hat M = ({\hat x}^1,\,...\,,{\hat x}^n)$ the following ordered form

\begin{equation}
\label{qz2-11}%
:{M}_{N_1...N_n}:\,=\,(\hat x^1)^{N_1}(\hat x^2)^{N_2}\, ...\,(\hat
x^n)^{N_n}
\end{equation}
where $N_i$ are natural numbers. In that case a 'noncommutativity degree' of
${\hat x}^i$ and ${\hat x}^j$ ($j>i$) will be measuring by the
following series in Plank constant

\begin{equation}
\label{qz2-12}%
\hat x^j\hat x^i\,=\,\hat x^i\hat x^j+\sum _{k=1}\hbar ^k
:f^{ij}_k:,\quad i<j,\quad  :f^{ij}_k:\in {\cal F}(\hat M) %
\end{equation}
where $:f^{ij}_k:$ are sums of ordered monomials (\ref{qz2-11}) with arbitrary
number of multipliers

\begin{equation}
\label{qz2-13}%
 :f^{ij}_k:\,=\,\sum _{N_1,\, ...\, ,N_n \geq 0;\,n\geq 0}
 f_k^{ij;N_1...N_n}:M_{N_1...N_n}: %
\end{equation}
and $f_k^{ij;N_1...N_n}$ are complex coefficients.

To introduce dynamics in the algebra of quantum observables ${\cal F}(\hat M)$ assume that evolution equations for $\hat x^i$ have the same form as for the commutative case

\begin{equation}
\label{qz2-17j}%
\frac{{\rm d}{\hat x}^i}{{\rm d}t} = \hat{X}(\hat x^i)
\end{equation}
but with a 'quantum' derivation $\hat X$ which includes quantum corrections. This derivation  is defined by its action on the basic elements

\begin{equation}
\label{qz2-14}%
\hat{X}(\hat x^i)\,=\,:{\hat X}^i: + \sum _{k=1}\hbar ^k
:g^i_k:\,,\quad :g^i_k:\in {\cal F}(\hat M).
\end{equation}
In the equation (\ref{qz2-14}) $:{\hat X}^i:$ means the ordered
functions ${\tilde X}^i(\tilde x)$ from the equation (\ref{qz2-7})
where all $\tilde x^i$ are replaced by $\hat x^i$ and $:g^i_k:$ can be
defined as

\begin{equation}
\label{qz2-15}%
:g^i_k:\, =\,\sum _{N_1,\, ...\, ,N_n \geq 0;\,n\geq 0}
g^{i;N_1...N_n}_k:M_{N_1...N_n}:,\quad%
g^{i;N_1...N_n} \in {\mathbb{C}}.%
\end{equation}

The commutation relations (\ref{qz2-12}) and the quantum derivation
(\ref{qz2-14}) include unknown coefficients $f_{k}^{ij;N_1...N_n}$
and $g_{k}^{i;N_1...N_n}$ which should be defined. To obtain them we
use time independence of the relations (\ref{qz2-12}). In view of the dynamical
equation (\ref{qz2-17j}) it means that the following relations hold

\begin{equation}
\label{qz2-20}%
\hat{X}(\hat x^j\hat x^i-\hat x^i\hat x^j-\sum _{k=1}\hbar ^k
:f^{ij}_k:)=0,\quad i<j,\quad i,j = 1, ...,n.
\end{equation}
Using the properties of a derivation and the formulas (\ref{qz2-12})
and (\ref{qz2-14}) one can obtain recurrent relations connecting
unknown coefficients $f_{k}^{ij;N_1...N_n}$ and
$g_{k}^{i;N_1...N_n}$.
The additional conditions for receiving these coefficients gives us
invariance of the commutation relations (\ref{qz2-12}) under
involution and Hermitian property of basics elements inherited from
(\ref{qz2-7m})

\begin{equation}
\label{qz2-21}%
(\hat x^i)^* =\hat x^i.
\end{equation}

Taking into account the explicit expression for quantum derivation
(\ref{qz2-14}) and definition of quantum integrals of motion

\begin{equation}
\label{qz2-19}%
\frac{{\rm d}\hat I}{{\rm d}t} = \hat X(\hat I) = 0
\end{equation}
one can obtain these integrals (if they exist) in the form

\begin{equation}
\label{qz2-18}
\begin{tabular}{l}
$
\hat I = :\hat I_0: + \sum _{k=1}\hbar ^k :\hat I_{k}:,
$
\\[12pt]
$
\quad :\hat I_{k}: = \sum _{N_1,\, ...\, ,N_n \geq 0;\,n\geq 0}I_{k}^{N_1...N_n}:\hat M_{N_1...N_n}:, \quad
I_{k}^{N_1...N_n} \in {\mathbb{C}}.
$
\end{tabular}
\end{equation}
In the above formula $:\hat I_0:$ is the ordered ${\tilde I}$  where all
$\tilde x^i$ are replaced by $\hat x^i$.

Thus, we have described the proposed quantization algorithm which, in principle, gives us an algebra of quantum observables with commutation relations (\ref{qz2-12}) among its basis elements, a quantum derivation (\ref{qz2-14}) with related quantum equations of motion (\ref{qz2-17j}), and a set of quantum integrals of motion (\ref{qz2-18}).
Now, let us turn to the examples which  illustrate the presented quantization method and allow us to clarify some its details.

\section{Examples}

In this section we apply the proposed method to quantization of some
low dimensional classical systems. These systems were chosen as examples
in spite of the fact that most of them are well known. This is because each of the systems has interesting specific features and it is important to demonstrate as the method deal with them. In what follows we omit the formal intermediate step (transition to commutative algebras) and simple calculations.

\subsection{Charged spinless particle in a magnetic field}

This example demonstrate that one can easily  obtain the standard formula  by the presented algorithm using only a measurable magnetic field  without references to a vector - potential (see, for instance, \cite{LL}, \cite{Bal}).

A charged spinless particle motion in a uniform magnetic field describes by the Newtonian equation

\begin{equation}
\label{qz3-1-0}%
m\frac{{\rm d}{\bf v}}{{\rm d}t} = \frac{q}{c}\,{\bf v}\times{\bf B}
\end{equation}
where $q$, $m$  and ${\bf v}$ are  particle's charge, mass and
velocity respectively, $c$ is the speed of light, ${\bf B}$ is a
magnetic field. For a plane motion  (when ${\bf v}$ is orthogonal to
${\bf B}$) the equation (\ref{qz3-1-0}) gives

\begin{equation}
\label{qz3-1-1}%
\frac{{\rm d}v_x}{{\rm d}t} = \frac{qB}{mc}v_y = X_m(v_x), %
\quad \frac{{\rm d}v_y}{{\rm d}t} = -\frac{qB}{mc}v_x = X_m(v_y)
\end{equation}
where $X_m$ is the related vector field. These equations up to notations and  interpretation of the dynamical variables coincide with the harmonic oscillator equations. So it is naturally expect that our quantization procedure will lead to the quantum oscillator algebra.
In the case under consideration  two velocity components $v_x,\, v_y$ play the role of the main dynamical variables. Any smooth function of the particle velocity $v$ can be considered as an integral of motion. We choose here the kinetic energy

$$ H = \frac{mv^2}{2}, \quad v^2=v_x^2+ v_y^2.$$

Turning to the quantization, let an ordered form of quantum monomials (\ref{qz2-11}) be $v_x^{N_1}v_y^{N_2}$. Then using the consistency condition (\ref{qz2-20}) we obtain the following simplest commutation relation

\begin{equation}
\label{qz3-1-4}%
\hat v_y\hat v_x = \hat v_x\hat v_y + {\rm i}\hbar k %
\end{equation}
where $k$ is an arbitrary real number. The quantum derivation

$$\hat  X_m(\hat v_x) = \frac{qB}{mc}\hat v_y,\,\,\quad \hat X_m(\hat v_y) =
-\frac{qB}{mc}\hat v_x$$ %
and the quantum kinetic energy (that is an integral of motion)
$$\hat H = \frac{m\hat v^2}{2}$$
remain without quantum corrections. Let us stress here that the condition (\ref{qz2-20}) does not give an unique solution. In the considered case instead a real number $k$ can be stood an arbitrary function of $\hat H$. To fix this function it is necessary an additional condition. Let the quantum derivation $\hat  X_m$ of the algebra ${\cal F}(\hat v_x,\hat v_y)$ be an inner one

\begin{equation}
\label{qz3-1-10}%
\hat  X_m(.) = \frac 1{\mathrm{i}\hbar }[\,.\,,\hat H],\quad \hat H\in {\cal F}(\hat v_x,\hat v_y).
\end{equation}
This condition means that the quantum equations of motion have the Heisenberg form $$
\frac{\mathrm{d}\hat v_x}{\mathrm{d}t}=
\frac 1{\mathrm{i}\hbar }[\hat v_x,\hat H],\qquad
\frac{\mathrm{d}\hat v_y}{\mathrm{d}t}=
\frac 1{\mathrm{i}\hbar }[\hat v_y,\hat H]
$$
and uniquely defines the constant $k=-\frac{qB}{c m^2}$.
Taking into account this value of $k$ we obtain from (\ref{qz3-1-4}) the well
known quantization rule for the velocity components

\begin{equation}
\label{qz3-1-5}%
[\hat v_x,\hat v_y] = {\rm i}\hbar\frac{qB}{m^2\,c}.
\end{equation}

\subsection{Euler top}

The example in this section shows that using our procedure one can obtain commutation relations of the angular momentum type directly from the classical equations without reference to their canonical representation by position and momentum operators.

The evolution of the angular momentum of a classical anisotropic top with
a fixed point at its center of inertia is defined by the Euler equations %

\begin{equation}
\label{qz3-3-1}
\begin{tabular}{l}
$
 \frac{\mathrm{d}L_1}{\mathrm{d}t}=a_1L_2L_3=X_{EU}(L_1)
$
\\[12pt]
$
\frac{\mathrm{d}L_2}{\mathrm{d}t}=a_2L_1L_3=X_{EU}(L_2)
$
\\[12pt]
$
 \frac{\mathrm{d}L_3}{\mathrm{d}t}=a_3L_1L_2=X_{EU}(L_3).
$
\end{tabular}
\end{equation}

where  $L_1,L_2,L_3$ are components of the angular momentum vector
and real parameters  $a_1,a_2,a_3$ depend on principal moments
of inertia $I_1,I_2,I_3$  by the formulas \cite{Arnold}

$$a_1={I_3}^{-1}-{I_2}^{-1},\quad a_2={I_1}^{-1}-{I_3}^{-1},\quad
a_3={I_2}^{-1}-{I_1}^{-1}.$$
It is convenient regard $a_1,a_2,a_3$ as components of a
three-dimensional vector ${\bf a} = (a_1,a_2,a_3).$ Quadratic
integrals of motion form a family

\begin{equation}
\label{qz3-3-3} H=\frac{1}{2}(b_1L_1^2+b_2L_2^2+b_3L_3^2)%
\end{equation}
where the parameters $b_1,b_2,b_3$ satisfy the relation

\begin{equation}
\label{qz3-3-4}
{\bf a}\cdot {\bf b} = a_1b_1+a_2b_2+a_3b_3 = 0%
\end{equation}
which means the orthogonality of the vectors ${\bf a}$  and ${\bf b}=(b_1,b_2,b_3)$.

The quantization algorithm gives in this case the following simplest commutation relations among the quantum dynamical variables

\begin{equation}
\label{qz3-3-5}%
\hat L_2\hat L_1 = \hat L_1\hat L_2 - \im\hbar f_3\hat L_3, \,\,
\hat L_3\hat L_1 = \hat L_1\hat L_3 + \im\hbar f_2\hat L_2, \,\,
\hat L_3\hat L_2 = \hat L_2\hat L_3 - \im\hbar f_1\hat L_1%
\end{equation}
and the quantum derivation with the following corrections


\begin{equation}
\label{qz3-3-6}
\begin{tabular}{l}
$
\hat X_{EU}(\hat L_1) = a_1\hat L_2\hat L_3 - \frac{\im\hbar a_1f_1}{2}\hat L_1
$
\\[12pt]
$
\hat X_{EU}(\hat L_2) = a_2\hat L_1\hat L_3 + \frac{\im\hbar a_2f_2}{2}\hat L_2
$
\\[12pt]
$
\hat X_{EU}(\hat L_3) = a_3\hat L_1\hat L_2 - \frac{\im\hbar a_3f_3}{2}\hat L_3.
$
\end{tabular}
\end{equation}
The real parameters ${\bf f}=(f_1,f_2,f_3)$ as well satisfy the  orthogonality condition

\begin{equation}
\label{qz3-3-7}%
{\bf a}\cdot {\bf f} = 0.%
\end{equation}
Note that the Euler quantum dynamical equations in view of (\ref{qz3-3-6}) can be rewritten in the symmetrized form

\begin{equation}
\label{qz3-3-8}
\begin{tabular}{l}
$
\frac{\mathrm{d}\hat L_1}{\mathrm{d}t}=\frac{a_1}{2}(\hat L_2\hat
L_3+\hat L_3\hat L_2)
$
\\[12pt]
$
\frac{\mathrm{d}\hat L_2}{\mathrm{d}t}=\frac{a_2}{2}(\hat L_1\hat
L_3+\hat L_3\hat L_1)
$
\\[12pt]
$
\frac{\mathrm{d}\hat L_3}{\mathrm{d}t}=\frac{a_3}{2}(\hat L_1\hat L_2+\hat L_2\hat L_1).
$
\end{tabular}
\end{equation}
As in the classical case the family of the quantum observables

\begin{equation}
\label{qz3-3-9}%
\hat H=\frac{1}{2}(b_1\hat L_1^2+b_2\hat L_2^2+b_3\hat L_3^2)%
\end{equation}
are the family of integrals of motion if the condition (\ref{qz3-3-4}) holds.
If we choose ${\bf b}$ such that the following relation satisfy

\begin{equation}
\label{qz3-3-10}%
{\bf b}\times {\bf f} = {\bf a}%
\end{equation}
where $\times $ denote a vector product in ${\mathbb{R}}^3$,
then related integral of motion  from the family (\ref{qz3-3-9}) generates the quantum equations of motion (\ref{qz3-3-8}) in the Heisenberg form.
We can choose from the family (\ref{qz3-3-9}) two independent
integrals of motion $\hat H,\,\hat H^{\prime}$ which are defined by
the vectors ${\bf b},\,{\bf b}^{\prime}$ and satisfy
(\ref{qz3-3-4}). These integrals are commutative when the following
condition is fulfilled

\begin{equation}
\label{qz3-3-11}%
 {\bf f}\cdot ({\bf b}\times {\bf b}^{\prime}) = 0.%
\end{equation}
In this case it is not difficult exercise  to obtain Fock space of common states for commuting $\hat H$ and $\hat H^{\prime}$ using, for example, Schwinger oscillator
representation of angular momentum type operators.

\subsection{Fourth order Pais - Uhlenbeck oscillator}
The example in this section shows that our procedure allows to quantize a system in a reasonable way whereas its standard quantization leads to the Hilbert space containing negative norm states \cite{PU},\cite{ManDav}.
The Pais - Uhlenbeck oscillator was introduced in the paper \cite{PU} as a toy model to study field theories with higher derivative terms. The details of the following calculations see in \cite{DS}.

The considered oscillator is defined by the fourth order equation

\begin{equation}
\label{qz3-2-1}%
\frac{\mathrm{d}^4x}{\mathrm{d}t^4} + ({\omega }_1^2+{\omega }_2^2)\frac{%
\mathrm{d}^2x}{\mathrm{d}t^2} + {\omega }_1^2{\omega }_2^2x=0.
\end{equation}
where $\omega _1,\ne \omega _2$ are positive frequencies. It can be
written in the form of a system of first order equations

\begin{equation}
\label{qz3-2-2}%
\begin{tabular}{l}
$\frac{\mathrm{d}x_1}{\mathrm{d}t} = x_2 = X_{PU}(x_1),\quad%
\frac{\mathrm{d}x_2}{\mathrm{d}t} = x_3 = X_{PU}(x_2)$%
\\[12pt]
$\frac{\mathrm{d}x_3}{\mathrm{d}t} = x_4 = X_{PU}(x_3),\quad%
\frac{\mathrm{d}x_4}{\mathrm{d}t} = -({\omega}_1^2 +{\omega}_2^2)x_3 - {\omega}%
_1^2 {\omega}_2^2x_1 = X_{PU}(x_4)$
\end{tabular}
\end{equation}
where $x_1=x $ is the coordinate in the original space, $x_2$ is the
velocity and so on. This system has two independent integrals of
motion

\begin{equation}
\label{qz3-2-3}
\begin{tabular}{l}
$H_1 = -\frac 12{\omega }_1^2{\omega }_2^2x_2^2 +\frac 12({\omega
}_1^2+{\omega }_2^2)x_3^2 +\frac 12x_4^2 +{\omega }_1^2{\omega
}_2^2x_1x_3, \quad $
\\[12pt]
$H_2= \frac 12{\omega }_1^2{\omega }_2^2x_1^2 +\frac 12({\omega
}_1^2+{\omega }_2^2)x_2^2 -\frac 12x_3^2 +x_2x_4,$
\end{tabular}
\end{equation}
Our quantization algorithm gives for quantum basis observables the following nonzero commutation relations

\begin{equation}
\label{qz3-2-4}%
\begin{tabular}{l}
$[\hat x_1,\hat x_2]=\mathrm{i}\hbar f, \quad [\hat x_1,\hat
x_4]=\mathrm{i}\hbar g,$ \\[6pt]
$[\hat x_2,\hat x_3]=-\mathrm{i}\hbar g, \quad [\hat x_3,\hat
x_4]=-\mathrm{i}\hbar {\omega }_1^2{\omega }_2^2f -\mathrm{i}\hbar ({%
\omega }_1^2+{\omega }_2^2)g$%
\end{tabular}
\end{equation}
where $f,g$ are arbitrary parameters. The dynamical equations  and
integrals of motion have no quantum corrections. Put a Hamiltonian in the form

$$\hat H = (f\omega _1^2+g)^{-1}(f\omega _2^2+g)^{-1}
(-f\hat H_1+g\hat H_2)$$%
where $\hat H_1,\hat H_2$ are $H_1,H_2$ with all $x_i$ are changed
by $\hat x_i$. Then the quantum dynamical equations can be written
in the Heisenberg form. The appropriate choice
of the parameters $f,g$ leads to the quantization of the Pais -
Uhlenbeck oscillator as the usual anisotropic oscillator with
positive norm states \cite{DS}.

\subsection{Nonlinear oscillator with coordinate - dependent mass}

The example in this section shows that our quantization algorithm allows to quantize  nonlinear systems. Here we consider an oscillator with coordinate - dependent mass (see \cite{CRS} and references therein). The evolution of this oscillator are defined by the following nonlinear equation

\begin{equation}%
\label{qz3-4-1}%
(1+\lambda x^2)\frac{\mathrm{d}^2x}{\mathrm{d}t^2}- \lambda
x(\frac{\mathrm{d}x}{\mathrm{d}t})^2 + \omega ^2x = 0,\quad \lambda \geq 0
\end{equation}%
and it has the non-polynomial integral of motion

\begin{equation}%
\label{qz3-4-4}%
H = \frac{1}{2}\frac{y^2+\omega ^2 x^2}{1+\lambda x^2}.
\end{equation}%
The equation (\ref{qz3-4-1}) is equivalent to the system

\begin{equation}%
\label{qz3-4-2}%
\frac{\mathrm{d}x}{\mathrm{d}t} = y,\quad \frac{\mathrm{d}y}{\mathrm{d}t} = \frac{\lambda
x}{1+\lambda x^2}y^2 - \frac{\omega ^2 x}{1+\lambda x^2}.
\end{equation}%
Using the quantization procedure we obtain the following simplest commutation relation

\begin{equation}%
\label{qz3-4-5}%
[\hat y,\hat x] =  -{\rm i}\hbar (1+\lambda \hat x^2)
\end{equation}%
quantum dynamical equations with quantum corrections

\begin{equation}%
\label{qz3-4-7}%
\frac{\mathrm{d}\hat x}{\mathrm{d}t} = \hat y,\quad%
\frac{\mathrm{d}\hat y}{\mathrm{d}t} =
\frac{\lambda \hat x}{1+\lambda {\hat x}^2}{\hat y}^2 -
\frac{\omega ^2 \hat x}{1+\lambda {\hat x}^2}
+ \ri\hbar\lambda (1 -\frac{2}{1+\lambda{\hat x}^2})\hat y
+ \frac{2\lambda ^2\hbar ^2\hat x}{1+\lambda {\hat x}^2}
\end{equation}%
and corrected quantum integral of motion

\begin{equation}%
\label{qz3-4-6}%
\hat H =\frac{1}{2}\frac{1}{1+\lambda{\hat x}^2}({\hat y}^2 + \omega ^2x^2)%
+ \frac{\ri\hbar\lambda\hat x}{1+\lambda{\hat x}^2}\hat y%
+ \hbar ^2\lambda (\frac{1}{1+\lambda {\hat x}^2 } - \frac{1}{2}).
\end{equation}%
It is easy to check that  $\hat H$ considered as a Hamiltonian allows to rewrite the equations (\ref{qz3-4-7}) in the Heisenberg form if one take into account the following formula

$$
[\hat y,\frac{1}{1+\lambda \hat x^2}] = \frac{2{\rm i}\hbar\lambda \hat x}{1+\lambda \hat
x^2}.
$$
This formula can be proof using the  integral representation

$$\frac{1}{1+\lambda \hat x^2} = \int _0^{\infty}
e^{-p(1+\lambda \hat x^2)} {\rm d}p$$
and well known decomposition

$$e^A\,B\,e^{-A} = B + [A,B] + \frac{1}{2}[A,[A,B]] + ...  \,\,.$$
The representation of the obtained algebra of quantum observables with commutation relation (\ref{qz3-4-5}) see in \cite{CRS}.

\vspace{10mm}
We conclude this work by some short remarks. The adduced examples were shown that the proposed quantization method is effective and effortlessly gives results in the simple cases. It suitable for multi-Hamiltonian systems (the examples 2) and 3)).
For Euler top the method gives the commutation relations of angular momentum type (\ref{qz3-3-5}) without usual handling of classical expressions. In the case of
Pais - Uhlenbeck oscillator our algorithm naturally leads to the positive norm Fock space of states. The example 4) was shown that the procedure works effectively for nonlinear systems as well. On the other hand, it follows from the examples that to obtain the unique result of quantization it is often necessary an additional condition besides the ones were declared in Section 2. In the considered cases it was the condition of Heisenberg form of quantum dynamical equations.

The presented examples do not cover most of interesting systems such as dissipative systems, system in a curved space-time, etc. We hope that quantization of these systems by the proposed method and  more precise mathematical formulation of the method itself will be carried out in the future.

\end{document}